\documentclass[aps,prd,floatfix,nofootinbib]{revtex4} 
\usepackage{graphics}
\usepackage{graphicx}

\usepackage{latexsym}
\usepackage[cp866]{inputenc}
\usepackage[english]{babel}  %
\input epsf

\begin{document}

\title{\boldmath Estimates of Absolute Branching Fractions for the $f_0(1710)$
Decays and Radiative Transitions $\psi(2S)\to\gamma f_0(1710)$ and
$\Upsilon(1S)\to\gamma f_0(1710)$}
\author{N. N. Achasov\,\footnote{achasov@math.nsc.ru}
and G. N. Shestakov\,\footnote{shestako@math.nsc.ru}}
\affiliation{\vspace{0.2cm} Laboratory of Theoretical Physics, S. L.
Sobolev Institute for Mathematics, 630090, Novosibirsk, Russia}


\begin{abstract}
Using the result of the VES Collaboration for $Br(J/\psi\to\gamma
f_0(1710))$, we estimate the absolute branching fractions for the
$f_0(1710)$ decays into $\pi\pi$, $K\bar K$, $\eta\eta$ and for the
first time into $\omega\omega$ and $\omega\phi$. In addition, we
estimate for the first time $Br(\psi(2S)\to\gamma f_0(1710))
\approx3.5\times10^{-5}$ and $Br(\Upsilon(1S)\to\gamma f_0(1710))
\approx1\times10^{-5}$.
\end{abstract}

\maketitle

Recently, the charge-exchange reaction $\pi^-p\to\omega\phi n$ was
studied with the upgraded VES facility (at the Protvino accelerator)
in the interaction of a 29 GeV pion beam with a beryllium target.
\cite{Do23}. The analysis performed showed that the observed signal
in $\omega\phi$ system can be described by the contribution of the
known scalar resonance $f_0(1710)$ \cite{PDG23}. The dominant
mechanism of the $\pi^-p\to f_0(1710)n$ reaction at high energies
and small momentum transfers is the one-pion exchange mechanism.
This fact allowed the authors of \cite{Do23} to find the product of
the $f_0(1710)\to\pi\pi$ and $f_0(1710)\to\omega \phi$ branching
fractions:
\begin{eqnarray}\label{Eq1}
Br(f_0(1710)\to\pi\pi)Br(f_0(1710)\to\omega\phi)=(4.8\pm1.2)\times
10^{-3}.\end{eqnarray} Then, using the data presented in the Review
of Particle Physics (RPP) \cite{PDG23} for the decays $J/\psi\to
\gamma f_0(1710)\to\gamma\pi\pi$ and $J/\psi\to\gamma f_0(1710)\to
\gamma\omega\phi$, they found the product $Br(J/\psi \to\gamma f_0
(1710)\to\gamma\pi\pi)Br(J/\psi\to\gamma f_0 (1710)\to\gamma\omega
\phi)=(9.5\pm2.6)\times10^{-8}$, divided it by Eq. (\ref{Eq1}),
extracted the root from this ratio and thus get the total branching
fraction of the $J/\psi\to\gamma f_0(1710)$ decay \cite{Do23}:
\begin{eqnarray}\label{Eq2}
\left[\frac{Br(J/\psi\to\gamma f_0(1710)\to\gamma\pi\pi) Br(J/\psi
\to\gamma f_0(1710)\to\omega\phi)}{Br(f_0(1710)\to \pi\pi)Br(f_0
(1710)\to\omega\phi)}\right]^{1/2}=\nonumber \\ =Br(J/\psi\to\gamma
f_0(1710))= (4.46 \pm0.82)\times10^{-3}.\qquad\qquad\end{eqnarray}
They compared this value with the known branching fraction for
$J/\psi\to \gamma f_0(1710)\to5\ channels$, i.e., with
$Br(J/\psi\to\gamma f_0 (1710)\to(\gamma\pi\pi+\gamma K\bar
K+\gamma\eta\eta+\gamma \omega
\omega+\gamma\omega\phi))=(2.13\pm0.18)\times10^{-3}$ \cite{PDG23},
and concluded that unregistered channels account for $Br(J/\psi\to
\gamma f_0(1710)\to(\gamma4\pi+\gamma\eta\eta'+\gamma\pi\pi K\bar K
+\cdot\cdot\cdot))=(2.33\pm0.84)\times10^{-3}$ \cite{Do23}.

If we now divide the values given in RPP \cite{PDG23} for the
branching fractions of the $J/\psi\to\gamma f_0(1710)\to\gamma
\pi\pi$, $\gamma K\bar K$, $\gamma\eta\eta$, $\gamma\omega\omega$,
$\gamma\omega\phi$ decays by $Br(J/\psi\to\gamma f_0(1710))$ from
Eq. (\ref{Eq2}), then we obtain the absolute branching fractions for
the corresponding decays of the $f_0(1710)$ resonance itself. These
estimates are presented in Table 1.\\

\begin{center}{\bf Table 1.} The branching fractions for the $f_0(1710)$
decays \vspace*{-0.2cm}\begin{table}  [!ht]
\begin{tabular}{ | c | c | c | c | c | c | }
  \hline
  \,\, $\pi\pi$\,\, & \,\, $K\bar K$\,\, & \,\, $\eta\eta$\,\, & \,\, $\omega\omega$\,\, & \,\, $\omega\phi$\,\, & \,\, Total \,\, \\ \hline
  \,\, $0.0852\pm0.0193$\,\, & \,\, $0.213\pm0.045$\,\, & \,\, $0.054\pm0.029$\,\, & \,\, $0.070\pm0.026$\,\, & \,\, $0.056\pm0.017$
  \,\, & \,\, $0.478\pm0.065$\,\, \\
  \hline
\end{tabular}\end{table}\end{center}

Note that the factorization of the effective creation and decay
coupling constants for the $f_0(1710)$ resonance makes it possible
to evaluate the ratio $Br(f_0\to\pi\pi)/ Br(f_0\to K\bar K)$ from
the data on the radiative $ J/\psi$, $\psi(2S)$, and $\Upsilon(1S)$
decays \cite{PDG23} in three different ways,
\begin{eqnarray}\label{Eq3} \frac{Br(f_0(1710)\to\pi\pi)}{Br(f_0
(1710)\to K\bar K)}=\frac{Br(J/\psi\to\gamma f_0(1710)\to\gamma
\pi\pi)}{Br(J/\psi\to\gamma f_0(1710)\to\gamma K\bar K)}=0.40\pm0.07,\\
\frac{Br(f_0(1710)\to\pi\pi)}{Br(f_0 (1710)\to K\bar K)}=\frac{Br(
\psi(2S)\to\gamma f_0(1710)\to\gamma \pi\pi)}{Br(\psi(2S)\to
\gamma f_0(1710)\to\gamma K\bar K)}=0.58\pm0.11,\\
\frac{Br(f_0(1710)\to\pi\pi)}{Br(f_0(1710)\to K\bar K)}=\frac{Br(
\Upsilon(1S)\to\gamma f_0(1710)\to\gamma \pi\pi)}{Br(
\Upsilon(1S)\to\gamma f_0(1710)\to\gamma K\bar K)}=0.36\pm0.17.
\end{eqnarray}
As is seen, these estimates are consistent with each other within
the error limits. The VES result (see Eq. (\ref{Eq2})),
factorization, and RPP data \cite{PDG23} allow us to determine the
absolute branching fractions for radiative decays $\psi(2S)\to\gamma
f_0(1710)$ and $\Upsilon(1S)\to\gamma f_0(1710)$ in two ways (using
the data on the $f_0(1710)\to\pi\pi$ and $f_0(1710)\to K\bar K$
channels):
\begin{eqnarray}\label{Eq4}
Br(\psi(2S)\to\gamma f_0(1710))=\frac{Br(\psi(2S)\to\gamma
f_0(1710)\to\gamma\pi\pi)}{Br(J/\psi\to\gamma f_0(1710)\to
\gamma\pi\pi)}Br(J/\psi\to\gamma f_0(1710))=(4.1\pm1.2)\times10^{-5},\\
Br(\psi(2S)\to\gamma f_0(1710))=\frac{Br(\psi(2S)\to\gamma
f_0(1710)\to\gamma K\bar K)}{Br(J/\psi\to\gamma f_0(1710)\to
\gamma K\bar K)}Br(J/\psi\to\gamma f_0(1710))=(3.1\pm0.7)\times10^{-5},\\
Br(\Upsilon(1S)\to\gamma f_0(1710))=\frac{Br(\Upsilon(1S)\to\gamma
f_0(1710)\to\gamma\pi\pi)}{Br(J/\psi\to\gamma f_0(1710)\to
\gamma\pi\pi)}Br(J/\psi\to\gamma f_0(1710))=(0.93\pm0.27)\times10^{-5},\\
Br(\Upsilon(1S)\to\gamma f_0(1710))=\frac{Br(\Upsilon(1S)\to\gamma
f_0(1710)\to\gamma K\bar K)}{Br(J/\psi\to\gamma f_0(1710)\to \gamma
K\bar K)}Br(J/\psi\to\gamma f_0(1710))=(1.03\pm0.19)\times10^{-5}.
\end{eqnarray}

The properties of the $f_0(1710)$ resonance and its possible nature
have been the subject of intense discussions for several decades,
see for review Refs. \cite{PDG23,KZ07} and references herein. For
now, the numerical estimates given in Table 1 are a good addition to
the very scarce information available about the absolute branching
fractions of the $f_0(1710)$ decays, see Section dedicated to this
state in RPP \cite{PDG23} (the existing data are not used by
Particle Data Group for finding averages, fits, limits, etc.).
Statements like ``seen'' in this Section can be superseded by the
corresponding values from Table 1.

A similar method for estimating the absolute branching fractions can
be useful for other heavy scalar (tensor) multichannel resonances,
for example, for $f_0(1370)$, $f_0(1500)$, $f_0(1770)$, and
$f_0(2020)$ \cite{PDG23} that can be produced both in $\pi N$
collisions via one-pion exchange mechanism and in radiative
$J/\psi$, $\psi(2S)$, and $\Upsilon(1S)$ decays.

In the recent work \cite{Ac23}, we discussed the properties of the
new $a_0(1700/1800)$ meson \cite{PDG23} assuming that it can be
similar to the $q^2\bar q^2$ state from the MIT bag \cite{Ja77}.
This $a_0$ meson has to have the isoscalar partner $f_0$ with a
close (or even degenerate) mass \cite{Ja77}. The branching fractions
in Table 1 will help us to understand whether the $f_0(1710)$ can
pretend to this role. This issue will be considered elsewhere.

The work was carried out within the framework of the state contract
of the Sobolev Institute of Mathematics, Project No. FWNF-2022-0021.


\end{document}